# Long-range optical trapping and binding of microparticles in hollow-core photonic crystal fibre

Dmitry S. Bykov[1], Shangran Xie[1,*], Richard Zeltner[1,2], Andrey Machnev[1], Gordon K. L. Wong[1], Tijmen G. Euser[1,3], and Philip St.J. Russell[1,2]

[1]*Max Planck Institute for the Science of Light, Staudtstr. 2, 91058 Erlangen, Germany*
[2]*Department of Physics, University of Erlangen-Nuremberg, 91058 Erlangen, Germany*
[3]*NanoPhotonics Centre, University of Cambridge, Cavendish Laboratory, CB3 0HE Cambridge, UK*
*\*shangran.xie@mpl.mpg.de*

Optically levitated micro- and nanoparticles offer an ideal playground for investigating photon-phonon interactions over macroscopic distances. Here we report the observation of long-range optical binding of multiple microparticles, mediated by intermodal scattering and interference inside the evacuated core of a hollow-core photonic crystal fibre (HC-PCF). Three polystyrene particles with 1 μm diameter are stably bound together with an inter-particle distance of ~40 μm, or 50 times longer than the wavelength of the trapping laser. The bound-particle array can be translated to-and-fro over centimetre distances along the fibre. When evacuated to 6 mbar gas pressure, the collective mechanical modes of the bound-particle array could be observed. The measured inter-particle distance at equilibrium and mechanical eigen-frequencies are supported by a novel analytical formalism modelling the dynamics of the binding process. The HC-PCF system offers a unique platform for investigating the rich optomechanical dynamics of arrays of levitated particles in a well-isolated and protected environment.

Since Ashkin's first report of the acceleration and trapping of microparticles by optical forces [1], optical tweezers has developed into a standard technique for biological manipulation and pico-Newton force sensing [2, 3], to mention just two from a wide range of applications. In recent years the emerging field of "levitated optomechanics" has attracted increasing interest. An optically tweezered particle, especially at low gas pressure, is isolated from the external environment, resulting in very low mechanical damping. This leads to very high mechanical Q-factors and permits particle rotational speeds in the MHz range [4, 5]. Recent advances include the use of feedback or cavity cooling of the centre-of-mass temperature towards the mechanical ground state [6-11], permitting fundamental tests of quantum mechanics [12-14], and the use of individual levitated particles as point sensors [15, 16]. Optical binding between arrays of trapped particles adds an additional dimension, and will result in rich dynamics, allowing access to collective coupling between high-Q mechanical oscillators and, potentially, simultaneous cooling of the mechanical motion of multiple particles.

Multiple trapping sites have previously been created using interference [17] and holographic tweezers [18], allowing formation of a lattice of trapped particles. In these experiments there is typically very little multiple scattering between particles – a necessary prerequisite for optical binding [19-21], which can only occur if the scattered field from one particle strongly interacts with the other particles in the array and vice-versa. In a 1D particle array, such as the one studied here, binding is possible because the optical fields propagate bidirectionally along the array.

The majority of 1D optical binding experiments to date have been performed in free space over distances that are limited by the Rayleigh range of the focusing optics. However, to control and measure the collective binding dynamics of such arrays, it is necessary to manipulate and monitor individual particles within the array without perturbing other degrees of freedom. This requires extended inter-particle distances and therefore long-range interactions between trapped particles.

In recent years, the optical binding range has been extended using non-diffracting Bessel beams [21] or by trapping particles in the evanescent field near the surface of a multimode glass microfibre [22]. These experiments were however conducted in a liquid environment, resulting in strongly damped collective dynamics. In addition, external spatial light modulators were used to create binding sites, limiting the stability and power handling of the system.

Here we report long-range optical binding of a chain of levitated particles inside the core of an evacuated HC-PCF. When the fundamental mode is launched into the HC-PCF, particle-induced scattering to higher-order guided modes results in an intermodal interference pattern and a landscape of periodically distributed trapping potentials within which subsequent particles can be trapped. The binding forces vary as the particles move relative to each other, and stable trapping and binding of a chain of particles results for configurations that locally minimize the free energy of the system. By offering modal fields that are tightly-confined over the entire fibre length, i.e., an infinite Rayleigh length, HC-PCF permits trapped particles to interact over distances orders of magnitude longer than the trapping wavelength, limited only by fibre and scattering loss. Absence of viscous damping at low gas pressure makes possible for the first time observation of the collective dynamics of a bound particle chain—an important first step towards "levitated collective optomechanics".

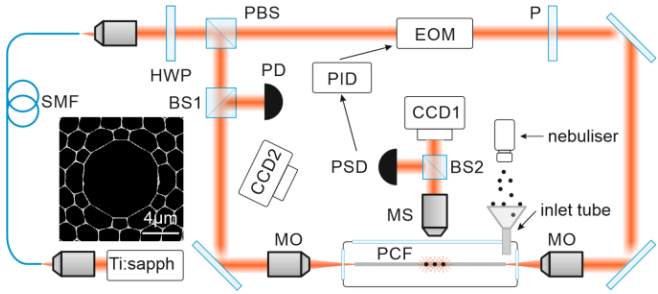

Fig. 1. Schematic of the experimental setup. SMF, single mode fiber; HWP, half-wave plate; PBS, polarizing beam splitter; MO, microscope objective; PD, photodiode; MS, microscope system; PSD, position sensitive detector; PID, proportional-integral-differential controller; EOM, electro-optical modulator; P, polarizer. Inset: scanning electron micrograph (SEM) of the HC-PCF.

The Experimental set-up is sketched in Fig. 1. Light from a pulsed Ti:sapphire laser (800 nm wavelength, 80 MHz repetition rate, 100 fs FWHM pulse duration) was delivered to the set-up through a 15 m length of single mode fiber (SMF, with group velocity dispersion $\beta_2 = 40$ fs/(m.THz)). This caused the pulses to broaden to FWHM durations of ~24 ps at the output of the SMF, with a frequency chirp of ~0.5 THz/ps. The pulses were split at a polarizing beam splitter (PBS) and coupled into opposite ends of an 8 cm length of HC-PCF with core diameter 7.8 µm, which was mounted inside a vacuum chamber (Fig. 1). A 96:4 beam splitter (BS1) and photodiode (PD) were used to monitor the transmitted power at one end of the HC-PCF. A medical nebuliser was used to inject polystyrene particles (typical diameters of 1 µm) into a dual-beam trap placed close to the fiber input face [23] (see SM1). Once trapped, a particle was propelled into the hollow core by adjusting the beam splitter so as to increase the forward power. Subsequent particles were launched into the fiber using the same technique. Once the desired number of particles had been launched, the chamber was sealed and pumped out to a pressure of a few mbar.

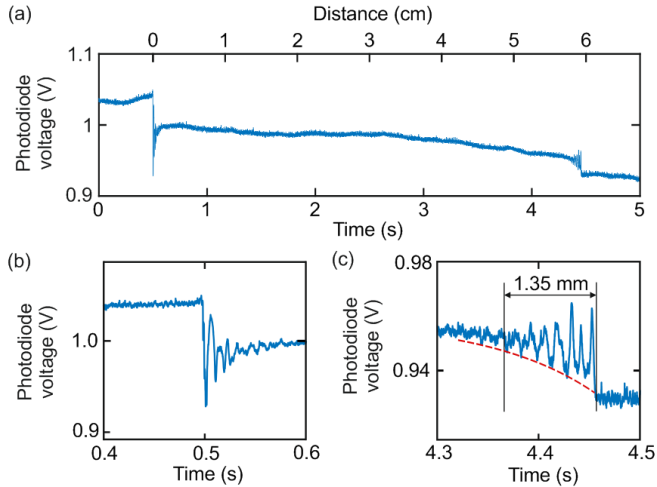

Fig. 2. (a) Temporal variation in transmitted power measured when the second particle enters the fiber. (b) Zoom into the time interval between 0.4 and 0.6 s, showing the fade-out of intermodal beating around fiber endface. (c) Zoom into the time interval from 4.3 to 4.5 s, when the second particle approaches the first. The red-dashed line is an exponential fit to the intermodal beating envelope.

When light is launched into the HC-PCF, it is difficult to avoid weak excitation of higher order modes (HOMs). Even a few percent of HOM power will cause quite strong intermodal interference, which however fades away with distance due to a combination of group velocity walk-off and pulse chirp. Thus, when a single particle is launched into the core, the transmitted power will fluctuate as the particle passes through the interference pattern, eventually stabilizing once the particle reaches a position where the stationary interference between the HOM pulses has faded away (Fig. 2(a) and (b)). The fade-away length can be written in the form (SM2):

$$z_F = \frac{\bar{v}_G \tau_0}{2} \sqrt{\frac{\bar{v}_G \pi \left(1 + (L/L_D)^2\right)}{\Delta v_G (L/L_D)}} \quad (1)$$

where $L_D = \beta_2/\tau_0$ is the dispersion length in the SMF, $\tau_0$ the 1/e HW pulse duration at the laser, $L$ the SMF length, $\Delta v_G$ the group velocity difference between the HC-PCF modes and $\bar{v}_G$ the mean group velocity (spectral broadening due to self-phase modulation in the SMF has been neglected, but is expected to be small). For the experimental parameters $z_F = 1.5$ mm.

A very similar effect is seen when two or more particles are close enough to sit in their respective intermodal interference patterns. Under these circumstances long-range optical binding can be observed. To explore these effects, we launched a 1-µm-diameter polystyrene particle into the fibre and followed this up 0.5 s later with a second similar particle. The power ratio was adjusted so that the second particle moved towards the first one. The transmitted power at the photodiode is plotted against time over a 5 second interval in Fig. 2(a). The effects of intermodal interference are seen at 0.5 s and 4.4 s, and zoom-ins of the responses in each case are shown in Figs. 2(b) and (c). As expected, intermodal interference causes the transmitted power to oscillate as the second particle moves away from the input face (Fig. 2(a)). After 0.55 s, the transmitted power becomes constant. Then at 4.36 s the first particle begins to be disturbed by intermodal interference created by the second particle (Fig. 2(c)). The oscillations increase in amplitude until the two particles become stably bound, when the modulation in the transmitted signal once again becomes constant. An exponential fit (dashed red curve) to the envelope of the oscillations yields a decay time of ~0.9 s which, given that the average speed of the particles is ~1.5 cm/s, corresponds to a fade-out distance of ~1.35 mm, in good agreement with the above analysis.

Then three particles were launched into the fiber in sequence, and after several attempts and adjustments it was found that they could be optically bound. Fig. 3(a) shows optical images of the bound-particle array at equilibrium, captured through the side of the HC-PCF using a CCD camera. The particles are spaced by 40±3 µm, the error being caused by the finite widths of the intensity peaks at the CCD.

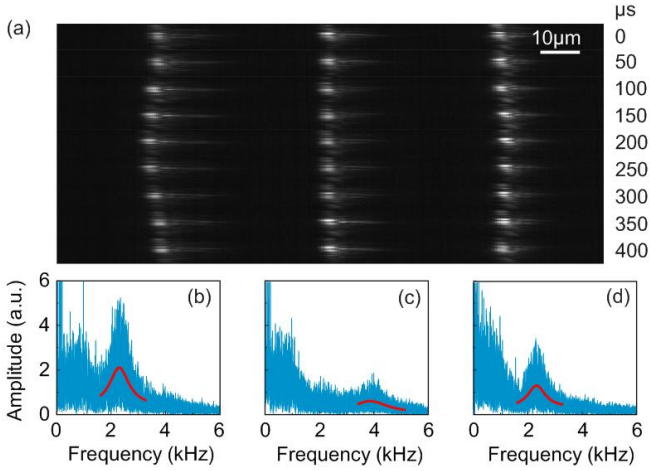

Fig. 3 (a) A series of snapshots of the bound-particle array during one period of the breathing mode, captured with a high-speed camera. (b) to (d): Spectra of the mechanical motion of the three bound particles. The blue curves are the measured data and the red curves are Lorentzian fits.

Thermally-driven vibrations of the bound-particle array (driven by Brownian motion) could be resolved at a pressure of 6 mbar, using a high speed video camera (20,000 frames per sec). Fig. 3(a) shows 9 consecutive frames at time intervals of 50 µs (see Media S2 for the video). During the measurement, electronic feedback was used to stabilize the particle array within the field of view of the imaging system using a position-sensing diode (PSD) to generate a signal proportional to the centre-of-mass motion of the particles. This was then fed back via a PID controller to the EOM.

The "breathing" mode of the bound particles, in which the two outer particles move out-of-phase while the central particle is stationary, can be directly observed in the video frames (Fig. 3(a)). The peak-to-peak amplitude of the vibrations is ~5 µm according to the scale bar in the video. In Figs. 3(b) to (d) the spectra of the thermally-driven centre-of-mass motion of particles 1, 2 and 3 are plotted, using data from the high speed video. The red-solid curves are Lorentzian fits to the individual spectral peaks. For particles 1 and 3 a strong peak occurs at 2.3 kHz (the frequency of the breathing mode), while for particle 2 a much weaker peak appears at 3.9 kHz, related to the case when the central particle moves out-of-phase with the two outer particles (see analysis below).

We now develop scattering matrix analysis to calculate the forces acting on each particle and thus identify configurations that result in stable binding and its dynamics. We represent the complex amplitudes of individual modes in the system by a column vector **v**, which is normalized so that $\mathbf{v} \cdot \mathbf{v}^* = 1$, meaning that the power in the $i$-th mode is $P_i = |v_i|^2 P_0$ where $P_0$ is the total power. The modes are assumed to form a complete orthogonal set.

Particle-induced scattering between incident and transmitted modes can then be described by $\mathbf{v}_{out} = [\mathbf{S}] \cdot \mathbf{v}_{in}$, where [**S**] is the scattering matrix and $\mathbf{v}_{in}$ and $\mathbf{v}_{out}$ are column vectors whose elements are the complex amplitudes of the incident and scattered modes. Orthogonality allows us to write the scattering coefficient from incident mode $i$ to forward-scattered mode $j$ in the form:

$$S_{ji} = \frac{\iint_A s_i m_j^* dA}{\iint_A |m_i|^2 dA} \quad (2)$$

where integration is over the transverse plane, $m_i(x, y)$ is the transverse field distribution of the $i$-th mode and $s_i(x, y)$ is the scattered field distribution immediately after the particle, which we calculate by 3D finite element modelling (FEM).

We now assume that back-scattering (which is very weak for the experimental parameters) and material absorption are negligible. We also assume that the particle is trapped at the centre of the core (this seems a good approximation since the launching beam is Gaussian) so that scattering is only among radially symmetric modes. Fig. 4(a) shows the geometry of the three-particle system.

The phase index, group index and loss of the modes were calculated by finite element modelling of the actual fiber microstructure, based on a high resolution scanning electron micrograph. Since linearly polarized $LP_{0i}$ modes of order $i > 3$ were found to have very high propagation loss, contributing negligibly to optical binding, we included just the first three lowest-order modes ($i = 1, 2, 3$), resulting in a 3×3 scattering matrix.

Since binding was observed at ~4 cm from the fiber input, and the fade-out distance of intermodal beating is ~1.35 mm, the three incident modes will each independently transfer momentum to the particles, because their instantaneous frequencies at each position along the fiber are different and no stationary intermodal beat-pattern can form. Also, since the pulse durations are orders of magnitude shorter than the mechanical response time of the particle (hundreds of µs), the particles will respond to the vector sum of the resulting optical momenta. Thus each incident mode can be treated independently.

For the $p$-th incident mode, the modal amplitudes on opposite sides of the particles are (using bra-ket notation to indicate left and right respectively):

$$\langle \mathbf{v}_1^{(p)}| = \mathbf{v}_{in}^{(p)}, \quad |\mathbf{v}_1^{(p)}\rangle = [\mathbf{S}]\langle \mathbf{v}_1^{(p)}| \quad (3)$$

for the first particle:

$$\langle \mathbf{v}_2^{(p)}| = [\mathbf{P}_{21}][\mathbf{S}]\mathbf{v}_{in}^{(p)}, \quad |\mathbf{v}_2^{(p)}\rangle = [\mathbf{S}]\langle \mathbf{v}_2^{(p)}| \quad (4)$$

for the second and :

$$\langle \mathbf{v}_3^{(p)}| = [\mathbf{P}_{32}][\mathbf{S}][\mathbf{P}_{21}][\mathbf{S}]\mathbf{v}_{in}^{(p)}, \quad |\mathbf{v}_3^{(p)}\rangle = [\mathbf{S}]\langle \mathbf{v}_3^{(p)}| \quad (5)$$

for the third. The propagation matrix $[\mathbf{P}_{kl}]$ between particle $l$ and particle $k$ is diagonal, with elements (1, $\exp[i\beta_{12}(z_k - z_l)]$, $\exp[i\beta_{12}(z_k - z_l)]$) where $\beta_{pq} = \beta_p - \beta_q$ is the propagation constant difference between $LP_{0p}$ and $LP_{0q}$ modes ($p < q$). The same analysis applies for the backward-propagating modes. The launched amplitudes of the three modes at the fiber input were estimated by calculating their overlap with a focused Gaussian beam (beam waist ~2.5 µm), yielding $\mathbf{v}_{in}^{(1)} = (0.982, 0, 0)$, $\mathbf{v}_{in}^{(2)} = (0, 0.143, 0)$ and $\mathbf{v}_{in}^{(3)} = (0, 0, 0.081)$.

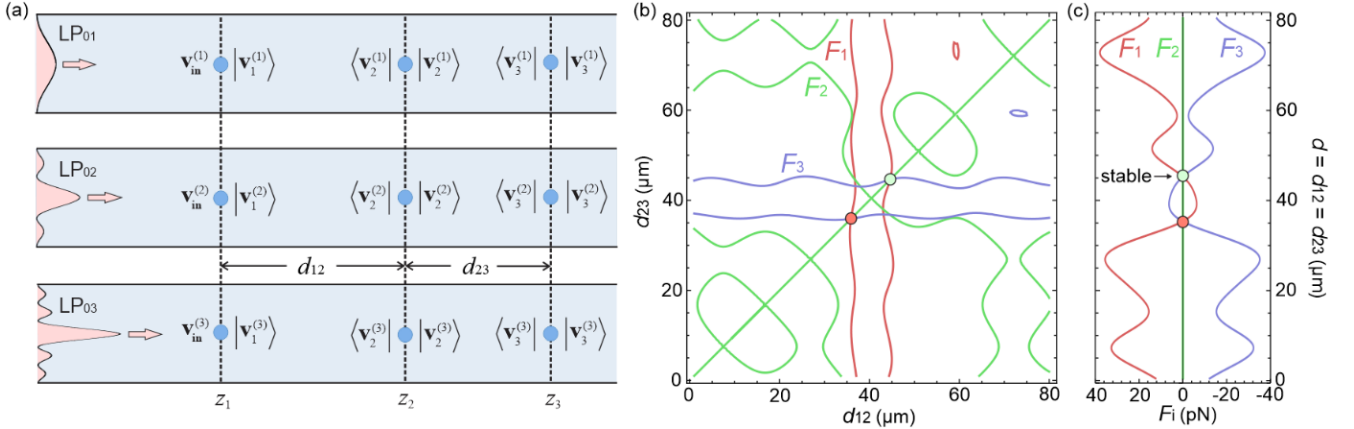

Fig. 4. (a) Sketch of the scattering matrix analysis, illustrating the notation used. Because the instantaneous frequencies of the three incident modes are different, no stationary intermodal fringes are seen. This means that the forces produced by each incident mode can be calculated independently. (b) Plots of the locii along which the forces on the particles are zero. Red: particle 1, green: particle 2 and blue: particle 3. Stable and unstable trapping points, when the forces on all three particles are zero, are marked by the green and red dots. (c) Optical forces acting on the three particles as a function of inter-particle distance d when the central particle is stationary.

To calculate the axial optical force $\vec{F}_k^p$ on the $k$-th particle, for incidence of the $p$-th mode on the first particle, we first need to find the local intensity on each side of the particle. This involves taking the modulus squared of the sum of the local field amplitudes. The total optical force on particle $k$, exerted by the forward-travelling light, is then given by:

$$\vec{F}_k = \frac{P_0}{2c} \sum_{p=1}^{3} \left( \left| \mathbf{N} \cdot \langle \mathbf{v}_k^{(p)} | \right|^2 - \left| \mathbf{N} \cdot | \mathbf{v}_k^{(p)} \rangle \right|^2 \right) \quad (6)$$

where $P_0/2$ is the total power incident in the forward direction (an equal amount is incident in the backward direction) and $c$ the speed of light *in vacuo*. The three-element row-matrix $\mathbf{N}$ has elements that correct for the overlap between the particle and the individual mode shapes:

$$N_i = \sqrt{\int_0^{a_p} J_0^2(u_{0i} r / a) 2\pi r dr \Big/ \int_0^{a} J_0^2(u_{0i} r / a) 2\pi r dr} \quad (7)$$

where $a$ is the core radius, $a_p$ the particle radius and $u_{0i}$ the $i$-th zero of the Bessel function $J_0$. For our experimental parameters, $N_1 = 0.250$, $N_2 = 0.362$ and $N_3 = 0.412$. The total optical force on particle $k$ is then found by adding the complementary force $\vec{\tilde{F}}_k$ exerted by the backward-propagating light (see SM3).

A necessary (but insufficient) condition for binding occurs when the optical forces acting on all three particles are simultaneously zero. Fig. 4(b) plots the lines of zero optical force on each particle as a function of the inter-particle distances $d_{12}$ and $d_{23}$. The diagonal line represents the symmetric configuration when $d_{12} = d_{23} = d$, and Fig. 4(c) plots the three forces versus $d$ in this case. All three forces are simultaneously zero at two positions, one of which is unstable (red dot), and the other stable (green dot). This predicts an inter-particle binding distance of $d = 44.6$ μm, in good agreement with the measurements. These binding positions are found to be relatively insensitive to the launched beam waist at each end of the fibre (in agreement with experiment), i.e., a change in beam waist from 2.5 μm to 2 μm results in the predicted inter-particle distance varying from 44.6 μm to 48.1 μm.

**Table 1: Frequencies and shapes of mechanical modes**

| $\Omega/2\pi$ (kHz) | $z_1$ | $z_2$ | $z_3$ |
|---|---|---|---|
| 3.85 | −0.30 | 0.91 | −0.30 |
| 2.24 | −0.71 | 0 | 0.71 |
| 0 | 0 | 0 | 0 |

The stiffness of the optical springs in each particle trap can be straightforwardly calculated by partial differentiation of the forces with respect to particle displacement. The result is a stiffness tensor $[\mathbf{K}]$ and the equation of motion $\ddot{\mathbf{z}}(t) + [\mathbf{M}]^{-1}[\mathbf{K}] \cdot \mathbf{z}(t) = \mathbf{0}$, where $[\mathbf{M}]$ is a diagonal matrix with elements given by the mass of each particle. For the experimental parameters ($P_0 = 50$ mW, particle mass $5.86 \times 10^{-16}$ kg):

$$[\mathbf{K}] = \begin{pmatrix} -0.10 & 0.08 & 0.016 \\ 0.13 & -0.26 & 0.13 \\ 0.016 & 0.08 & -0.10 \end{pmatrix} \text{pN/μm} \quad (8)$$

and the mechanical frequencies and shapes of collective eigenmodes can be calculated (Table 1), agreeing well with the observation.

In summary, multiple polystyrene microparticles can be optically bound by intermodal interference within the evacuated core of a HC-PCF where, protected from environmental disturbance, their collective vibrational modes can be resolved. Using electronic feedback to stabilise the mechanical motion at ultralow pressure, it should be possible to reach much higher mechanical Q-factors and explore cooling to the ground-state, nonlinear coupling and synchronization of the motion of multiple particles. By adjusting the trapping pulse chirp and duration and fibre dispersion, the binding length (40 μm in the experiments reported here) can be increased, potentially allowing particle binding over centimetre distances. Finally, the whole assembly of particles can be moved to-and-fro along the HC-PCF, suggesting applications in remote sensing.

# Long-range optical trapping and binding of microparticles in hollow-core photonic crystal fibre: supplementary information

Dmitry S. Bykov[1], Shangran Xie[1,*], Richard Zeltner[1,2], Andrey Machnev[1], Gordon K. L. Wong[1], Tijmen G. Euser[1,3], and Philip St.J. Russell[1,2]

[1]*Max Planck Institute for the Science of Light, Staudtstr. 2, 91058 Erlangen, Germany*
[2]*Department of Physics, University of Erlangen-Nuremberg, 91058 Erlangen, Germany*
[3]*NanoPhotonics Centre, University of Cambridge, Cavendish Laboratory, CB3 0HE Cambridge, UK*
*\*shangran.xie@mpl.mpg.de*

### S1. Procedure of trapping particles in front of the HC-PCF

The aerosol launching procedure described by Ref [1] was used to trap the particle in front of the HC-PCF: the particles were first dissolved in water in a particle-to-water mass ratio of $10^{-3}$–$10^{-2}$. A medical nebulizer was used to produce droplets of the particle solution, with each droplet containing a single particle on average. The droplets were injected into the chamber via an inlet tube in the lid that was placed above-in-front of the fibre until one of them was trapped in front of the HC-PCF core. The laser irradiation usually evaporates the droplet within a few seconds, leaving the particle in the dual-beam optical trap.

A trapping event can be observed either via dropping of the transmitted power (from fractions of percent for 100 nm particle up to 5% for 1 µm particle) or by observing changes in the image of the fibre endface taken by an external camera (CCD2 in Fig. 1a). As shown in Media S1, the image of the fibre endface starts to wobble a bit due to Brownian motion of the trapped particle. Bigger particles can be launched by using a glass plate mounted on a piezoelectric transducer which is placed below the fibre endface. The piezoelectric transducer can be driven at the resonance frequency of the glass plate, catapulting the particles in front of the fibre.

### S2. Analysis of pulse walk-off distance inside HC-PCF

In our experiment, the pulsed Ti:sapphire laser (80 MHz repetition rate, 100 fs FWHM pulse duration) was delivered to the set-up through a 15 m length of SMF. The in-coupling to the HC-PCF and the trapped particle would excite HOMs with different group velocities. The walk-off distance of the pulsed trapping beam inside HC-PCF therefore contributed from both the pulse chirping in SMF and the modal walk-off in HC-PCF. Assuming a Gaussian pulse from the initial Ti:sapphire laser with field amplitude written as

$$E(\tau) = \exp\left(-\tau^2 / (2\tau_0^2)\right) \quad (S1)$$

where $\tau$ is the time frame, $\tau_0$ is the 1/e half-width. The pulse envelop after propagating a distance $L$ along the SMF with group velocity dispersion $\beta_2$ is [2]

$$E(\tau, L) = \frac{\tau_0}{\sqrt{\tau_0^2 - i\beta_2 L}} \exp\left(\frac{-\tau^2}{2(\tau_0^2 - i\beta_2 L)}\right) \quad (S2)$$

with 1/e half-width given by

$$\tau_{1/2} = \tau_0 \sqrt{1 + \beta_2^2 L^2 / \tau_0^4} = \tau_0 \sqrt{1 + L^2 / L_D^2} \quad (S3)$$

where $L_D$ is the dispersion length. Given $\beta_2 = 40$ fs/(m.THz) for SMF and $L = 15$ m, the FWHM of the chirped pulse at the output end of the SMF is ~24 ps. The instantaneous frequency deviation from the carrier frequency $\omega_0$ is given by

$$\delta\omega(\tau, L) = \frac{2L/L_D}{1+(L/L_D)^2} \frac{\tau}{\tau_0^2} \quad (S4)$$

Inside HC-PCF, the excited $LP_{0i}$ modes will walk off from each other, with the walk-off time $\Delta\tau = z\Delta v_G / \bar{v}_G^2$, where $z$ is propagation distance inside HC-PCF, $\Delta v_G$ the group velocity difference between the modes and $v_G$ is the mean modal group velocity. Therefore the several modal pulses would have different frequencies, with the frequency offset taking the value:

$$\Delta f = \frac{2L/L_D}{1+(L/L_D)^2} \frac{z\Delta v_G}{\pi \bar{v}_G^2 \tau_0^2} \quad (S5)$$

Here we ignore the weak group velocity dispersion of HC-PCF compared to SMF. Intermodal beating will fade away mainly due to temporal oscillations that average the beating to zero. This will occur approximately when $\Delta f \cdot z / \bar{v}_G \leq 1/2$, which means that intermodal beating will fade out after a distance:

$$z_F = \frac{\bar{v}_G \tau_0}{2} \sqrt{\frac{\bar{v}_G \pi (1+(L/L_D)^2)}{\Delta v_G (L/L_D)}} \quad (S6)$$

which is identical to Eq. (1) in the primary manuscript.

### S3. Optical force induced by the backward-propagating light

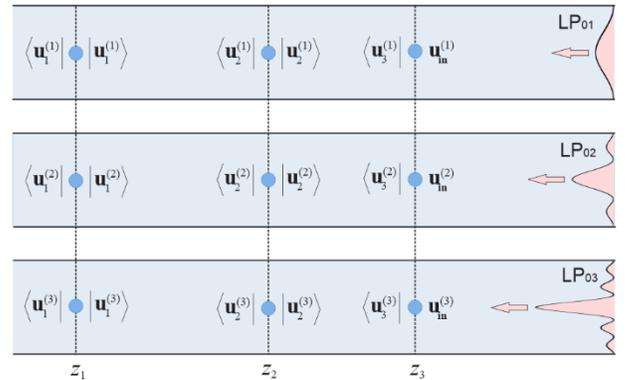

Fig. S1. Sketch of the scattering matrix analysis for the backward-propagating light.

Following the analysis of forward-propagating light in the primary manuscript, Fig. S1 sketches the configuration and notation for the backward-propagating light. Here we use the column vector **u** to represent the complex amplitudes of fibre modes. In this case, for the $p$-th incident mode, the modal amplitudes on opposite sides of the particles are:

$$\left|\mathbf{u}_3^{(p)}\right\rangle = \mathbf{u}_{in}^{(p)}, \quad \left\langle\mathbf{u}_3^{(p)}\right| = [\mathbf{S}]\left|\mathbf{u}_3^{(p)}\right\rangle \tag{S7}$$

for the third particle:

$$\left|\mathbf{u}_2^{(p)}\right\rangle = [\mathbf{P}_{23}][\mathbf{S}]\mathbf{u}_{in}^{(p)}, \quad \left\langle\mathbf{u}_2^{(p)}\right| = [\mathbf{S}]\left|\mathbf{u}_2^{(p)}\right\rangle \tag{S8}$$

for the second and :

$$\left|\mathbf{u}_3^{(p)}\right\rangle = [\mathbf{P}_{12}][\mathbf{S}][\mathbf{P}_{23}][\mathbf{S}]\mathbf{u}_{in}^{(p)}, \quad \left\langle\mathbf{u}_3^{(p)}\right| = [\mathbf{S}]\left|\mathbf{u}_3^{(p)}\right\rangle \tag{S9}$$

for the first. $[\mathbf{P}_{kl}]$ is defined in the primary manuscript, and $\mathbf{u}_{in}^{(p)} = \mathbf{v}_{in}^{(p)}$. The optical force contributed from the backward-propagating light shares the same form of Eq. (6) of the primary manuscript except replacing **v** with **u**:

$$\bar{F}_k = \frac{P_0}{2c}\sum_{p=1}^{3}\left(\left\|\mathbf{N}\cdot\left\langle\mathbf{u}_k^{(p)}\right\rangle\right\|^2 - \left|\mathbf{N}\cdot\left|\mathbf{u}_k^{(p)}\right\rangle\right|^2\right) \tag{S10}$$

The total optical force acting on particle $k$ is the sum of the forward and backward contributions, thus:

$$F_k = \vec{F}_k + \bar{F}_k \tag{S11}$$

**Media S1:** Launching multiple particles inside HC-PCF
**Media S2:** Breathing modes of the bound-particle array